\documentclass[aps,preprint,showpacs, showkeys]{revtex4}%
\usepackage{amsfonts}
\usepackage{amsmath}
\usepackage{amssymb}
\usepackage{graphicx}
\usepackage{epsfig}
\usepackage{exscale}
\usepackage{float}
\usepackage{bm}
\usepackage{suffix}
\usepackage{mathtools}
\usepackage{newunicodechar}
\usepackage{bbding}
\usepackage{tikz}
\usepackage{multirow}%
\usepackage[shortlabels]{enumitem}
\providecommand{\U}[1]{\protect\rule{.1in}{.1in}}

\DeclarePairedDelimiterX\MeijerM[3]{\lparen}{\rparen}
{\begin{smallmatrix}#1 \\ #2\end{smallmatrix}\delimsize\vert\,#3}
\newcommand\MeijerG[8][]{  G^{\,#2,#3}_{#4,#5}\MeijerM[#1]{#6}{#7}{#8}}
\WithSuffix
\newcommand\MeijerG*
[7]{
G^{\,#1,#2}_{#3,#4}\MeijerM*{#5}{#6}{#7}}

\begin{document}
\title[ ]{ Critical Exponents of the O(N)-symmetric $\phi^4$ Model from the \boldmath{\large{ $\varepsilon^7$}} Hypergeometric-Meijer Resummation}
\author{Abouzeid M. Shalaby}
\email{amshalab@qu.edu.qa}
\affiliation{Department of Mathematics, Statistics, and Physics, Qatar University, Al
Tarfa, Doha 2713, Qatar}

\begin{abstract}

We extract the $\varepsilon$-expansion from the recently obtained seven-loop $g$-expansion for the renormalization group functions of the $O(N)$-symmetric model. The different series obtained for the critical exponents $\nu,\ \omega$ and $\eta$ have been resummed using our recently introduced hypergeometric-Meijer resummation algorithm. In three dimensions,  very precise results have been obtained for all the critical exponents for $N=0,1,2,3$ and $4$. To shed light on the obvious  improvement of the predictions at this order, we obtained  the divergence of the specific heat critical exponent $\alpha$ for the $XY$ model. We found  the result $-0.0123(11)$ which is compatible with the famous experimental result of -0.0127(3) from the specific heat of zero gravity  liquid helium superfluid transition  while the   six-loop Borel with conformal mapping  resummation result in literature  gives the value -0.007(3). For the challenging case of resummation of the $\varepsilon$-expansion series in two dimensions, we showed that our resummation results reflect a significant improvement to the previous six-loop resummation predictions.
\end{abstract}
\maketitle
\section{Introduction}

Quantum field theory (QFT) offers a successful way to study critical phenomena
in many physical systems \cite{zinjustin,zin-borel,Berzin,Kleinert-Borel,kleinert,kleinert2,zin-cr,Eta4,Guillou,Kleinert5L,pelsito}. It is universality that is behind the scene where different systems sharing
the same symmetry properties follow the conjecture that    they ought to   behave in  a similar manner at phase transition. So it
is not strange to have a fluid possessing  the same critical exponents like a magnetic
one when both lie in the class of universality. The $O(N)$ vector model from
scalar field theory has an infra red attractive fixed point and possesses the
symmetry that can describe many physical  systems at phase transitions. Near phase transitions, the theory is totally non-perturbative 
where in literature there exist many computational trends used to study the critical phenomena within the $O(N)$ vector model.

For the study of critical phenomena in the $O(N)$ vector model, Monte Carlo
simulations have been used successfully and give precise results for the
critical exponents \cite{MC10,MC11,nuN0E,nuN0,MCN2,MC01,MC02,MC16,MC19,MC19a}. Besides,  bootstrapping  the model in three dimensions
has been accomplished recently and researchers succeeded to obtain precise results
\cite{Bstrab,Bstrab5,Bstrab2,Bstrab3,Bstrab4,BstrabN0}. The nonperturbative
renormalization group has been applied to the same model and gives accurate
results too \cite{NPRG}. Apart from these non-perturbative methods, the oldest
way to tackle the critical phenomena in QFT is resummation techniques applied
to resum the divergent perturbation series associated with renormalization group (RG) functions of that model. However, the precision of RG results has not been improved since 1998 \cite{zin-exp,dispute,dispute1} and thus in a need  to push it forward to a precision that make it able to compete  with recent accurate results from Monte Carlo simulations and bootstrap calculations.

The most traditional resummation algorithm is Borel and its extensions which have been widely used in literature
\cite{Berzin,Kleinert-Borel,zin-cr,Eta4,Guillou,zin-exp,ON17,Prd-GF}. In fact,  the
recent progress in  obtaining higher orders of the perturbation series stimulates the need for the application of the resummation techniques to investigate the theory. Regarding that, the six-loop  of the renormalization group functions has been recently obtained \cite{ON17} and then the seventh order   has been obtained too \cite{7L}. These orders are representing the renormalization group functions within
the minimal subtraction regularization scheme    in $D=4-\varepsilon$ dimensions. 

The study  of critical phenomena by finding an approximant to  the perturbation series follows different routes. For instance, perturbative calculations at fixed $D$ dimensions \cite{zin-exp,Kleinert-Borel} are always giving better results
specially in three dimensions. However, while exact results are known in two
dimensions, the resummation of perturbation series did not give reliable
results for some exponents \cite{Borelg2,sokolov}. This point has been studied
  in Refs. \cite{ON17,sokolov,plesito2} and it has been argued there that the reason behind this is thought to be the
non-analiticity of the $\beta-$function at the fixed point. For the $\varepsilon$-expansion on the other hand, perturbation series though possesses slower convergence  \cite{Kleinert-Borel,ON17},  might not  suffer from non-analiticity issues like the $g$-series \cite{analytic}. In view of the recent seven-loops ($g$-expansion) calculations \cite{7L}, one thus can aim to get improved results from  resumming the corresponding seventh order of $\varepsilon$ expansion in three dimensions ($\varepsilon=1$) as well as get improved  predictions  for two dimensions $\varepsilon=2$. A note to be mentioned here is that the most accurate renormalization group prediction for the exponent $\nu$ for the $XY$ model \cite{zin-exp} (for instant) has a relatively large uncertainty. This makes it  excluded   from playing a role in the current $\lambda$-point dispute \cite{dispute,dispute1}. Accordingly, higher order predictions from renormalization group is more than important.   

Inspired by the simple hypergeometric algorithm in Ref.\cite{Prl}, in  previous articles \cite{abo-expon,Abo-large}, we introduced and applied the hypergeometric-Meijer
resummation algorithm. What makes our algorithm preferable is its simplicity
and of having no arbitrary parameters like Borel algorithm and its extensions. Besides, it
gives very competitive predictions when compared to the more sophisticated 
Borel with conformal mapping algorithm for instance. The algorithm has been
applied successfully for the six-loop $\varepsilon-$series and for the
seven-loop coupling series in Ref.\cite{abo-expon}. For expansions in $4-\varepsilon$ dimensions, however, it is always believed that
the $\varepsilon-$series has better convergence than the coupling-series \cite{Kleinert-Borel}. In
fact one can speculate about this by considering the large order behavior for
both series. For the coupling series, the large order behavior includes the
term $\left(  -g_{c}\right)  ^{n}$ ( $g_c\equiv$ critical coupling) while the $\varepsilon-$series has the term
$\left(  -\sigma\varepsilon\right)  ^{n}$ with  $\sigma=\frac{3}{N+8}$. For
$N=1$ and in three dimensions ( for instance),  at the fixed point the $g$-series behaves as $\left(
-0.47947\right)  ^{n}$ ( from seven-loops calculations)  while    the
$\varepsilon-$series behaves as  $\left(  -0.33333\right)  ^{n}$. So it
is expected that the resummation of the $\varepsilon-$series has better convergence. 

The recent resummation results of the six-loop  $\varepsilon$-series \cite{ON17,abo-expon} gave accurate predictions for the critical exponents $\nu,\eta$ and $\omega$ for the $O(N)$-symmetric $\phi^4$ theory. However, the predictions of the relatively small exponents like divergence of specific heat exponent $\alpha$ are still far away from expected results. For the $XY$ model for instance, our hypergeometric-Meijer algorithm  gives the result $\alpha=-0.00886$ \cite{abo-expon} (from six loops) while Borel with conformal mapping result in Ref.\cite{ON17} is $-0.007(3)$ and the resummation of seven-loop $g$-series in Ref.\cite{abo-expon} predicts the value $-0.00860$. All of these predictions are all not close enough to the result of the famous experiment in Ref. \cite{alphaxy}. In that reference, the measurement  of the  specific heat of liquid helium in zero gravity  yields  the result $-0.0127(3)$. Moreover, either the six-loop $\varepsilon$-expansion or the seven-loop $g$-expansion are not giving results that overlap with  Monte Carlo and conformal bootstrap results \cite{dispute1}.  Accordingly, resumming the seven-loop $\varepsilon$-series represents an important point to monitor the improvement of the RG predictions of the critical exponents.  With that in mind,  our aim in this work is to first obtain the
$\varepsilon-$series corresponding to the recent seven-loop coupling series
for the $\beta$,$\gamma_{m^{2}}$ and $\gamma_{\phi}$ renormalization group
functions and then apply our resummation algorithm to the  series representing the 
critical exponents $\nu$, $\eta$ and $\omega$ for the $O(N)-$symmetric quantum
field model.

The organization of this paper is as follows. In Sec.\ref{algo}, a brief description of the hypergeometric-Meijer resummation algorithm is introduced. We present in Sec.\ref{7L-eps} the extracted seven-loop $\varepsilon$-expansion of the renormalization group functions. The resummation of the different $\varepsilon$-series  representing the critical exponents $\nu,\eta$ and $\omega$ is presented in Sec.\ref{resum7}. In this section a comparison with predictions from other methods for $N=0,1,2,3$ and $4$ is listed in different tables for each $N$ individually. The study of the challenging two-dimensional case will follow in Sec.\ref{2dim}. The last section in this paper (Sec.\ref{conc}) is dedicated for summary and conclusions.

\section{The hypergeometric-Meijer Resummation algorithm}\label{algo}

To make the work self consistent, we summarize in this section the
hypergeometric-Meijer resummation algorithm that was firstly introduced in
Ref.\cite{Abo-large} and then applied to the six-loop ($\varepsilon$-expansion) and seven-loop $g$-expansion in Ref.\cite{abo-expon}. However, the error analysis of the resummation process will be introduced also here. Accordingly, the algorithm will be presented in two subsections one for the resummation process and one for error calculations. 

\subsection{Hypergeometric-Meijer Resummation}\label{algo1}

Our hypergeometric-Meijer Resummation is a natural extension of  the simple hypergeometric approximants introduced by Mera \textit{et.al} in Ref.\cite{Prl}. The authors in that reference suggested the hypergeometric function   $_{2}F_{1}(a_{1},a_{2};b_{1};-\sigma x)$ as an approximant for  a divergent series with zero radius of convergence. However, it has been realized by the same authors that the suggested approximant has a series expansion with finite radius of convergence and thus clarified that the prediction is less accurate for small values of the perturbation parameter \cite{Prd-GF,cut}. Mera \textit{et.al} resolved this issue via the use of a Borel-Pad$\acute{e}$ technique for which the Borel functions are the hypergeometric functions $_{p}F_{p-1}(a_{1},a_{2},....,a_{p};b_{1},b_{2},....b_{p-1};x)$. In Refs.\cite{abo-expon,Abo-large}, we tried to resolve the same issue in a simpler way as well as in a way to have approximants that can employ the known parameters from the asymptotic behavior of the given divergent series. These parameters ( representing strong coupling and large-order asymptotic behaviors) are well known to accelerate the convergence of the resummation process \cite{zinjustin,Kleinert-Borel}. Our idea is based on selecting the hypergeometric approximants that possesses all the known features of the given perturbation series. We found that out of the hypergeometric approximants $_{p}F_{q}(a_{1},a_{2},....,a_{p};b_{1}%
,b_{2},....b_{p-2};-\sigma x)$, only the hypergeometric functions  $_{p}F_{p-2}(a_{1},a_{2},....,a_{p};b_{1}%
,b_{2},....b_{p-2};-\sigma x)$   are able to be parametrized to give the known weak-coupling information, the large-order asymptotic form and the strong-coupling behavior of divergent series  with an $n!$ growth factor. To elucidate the process more, consider a perturbation series of a physical quantity $Q$ for which the first $M+1$ terms are known:
\begin{equation}
Q\left(  x\right)  \approx\sum_{0}^{M}k_{i}x^{i}. \label{pertQ}%
\end{equation}
Assume that the asymptotic large-order  behavior for the series is also known to be of the
form: 
\begin{equation}
c_{n}=\alpha n!(-\sigma)^{n}n^{b}\left(  1+O\left(  \frac{1}{n}\right)
\right)  ,\text{ \ \ }n\rightarrow\infty. \label{large-order}%
\end{equation}
As shown in Ref.\cite{Abo-large}, the hypergeometric series $_{p}F_{p-2}%
(a_{1},a_{2},....,a_{p};b_{1},b_{2},....b_{p-2};-\sigma x)$ can reproduce the
same large-order behavior with constraint on its numerator and denominator
parameters as:%
\begin{equation}
\sum_{i=1}^{p}a_{i}-\sum_{i=1}^{p-2}b_{i}-2=b. \label{constr}%
\end{equation}
So the hypergeometric series $_{p}F_{p-2}(a_{1},a_{2},....,a_{p};b_{1}%
,b_{2},....b_{p-2};-\sigma x)$ \ possesses all the known features of the given
series when matching order by order the first $M+1$ coefficients from the
perturbation series in Eq.(\ref{pertQ}) with the first $M+1$ coefficients of
the expansion of the hypergeometric function $_{p}F_{p-2}(a_{1},a_{2}%
,....,a_{p};b_{1},b_{2},....b_{p-2};-\sigma x)$. This type of hypergeometric functions have the expansion:
\begin{equation}
_{\text{ }p}F_{p-2}\left(  {a_{1},......a_{p};b_{1},........b_{p-2};-\sigma
x}\right)  =\sum_{n=0}^{\infty}\frac{\frac{\Gamma\left(  a_{1}+n\right)
}{\Gamma\left(  a_{1}\right)  }....\frac{\Gamma\left(  a_{p}+n\right)
}{\Gamma\left(  a_{p}\right)  }}{n!\frac{\Gamma\left(  b_{1}+n\right)
}{\Gamma\left(  b_{1}\right)  }....\frac{\Gamma\left(  b_{p-2}+n\right)
}{\Gamma\left(  b_{p-2}\right)  }}\left(  -\sigma x\right)  ^{n}.
\end{equation}
Once parametrized by matching with the given series,  the divergent hypergeometric series is now known up to any order
and can be resummed by using its representation in terms of the Meijer-G
function of the form \cite{HTF}:
\begin{equation}
_{\text{ }p}F_{q}(a_{1},...a_{p};b_{1}....b_{q};x)=\frac{\prod_{k=1}^{q}%
\Gamma\left(  b_{k}\right)  }{\prod_{k=1}^{p}\Gamma\left(  a_{k}\right)  }%
\MeijerG*{1}{p}{p}{q+1}{1-a_{1}, \dots,1-a_{p}}{0,1-b_{1}, \dots, 1-b_{q}}{x}.
\label{hyp-G-C}%
\end{equation}
Note that for $M$ even, $M$ equations are generated by matching with the available orders from the given perturbation series to solve for $M=(2p-2)$ unknown parameters in the hypergeometric function. In the odd $M$ case,  we employ the constraint in Eq.(\ref{constr}) to get $M+1$ equations to solve for the  $M+1$ unknown parameters. In any case, we always need an even number of equations to determine the $2p-2$ unknown parameters.

To give an example, consider the lowest order approximant (two-loops)
$_{2}F_{0}(a_{1},a_{2};\ ;-\sigma x)$ when matched we get the results:
\begin{align}
-a_{1}a_{2}\sigma &  =k_{1}\nonumber\\
\frac{1}{2}a_{1}\left(  1+a_{1}\right)  a_{2}\left(  1+a_{2}\right)  \left(
-\sigma\right)  ^{2}  &  =k_{2}. \label{w-algo}%
\end{align}
 These equations are solved for the unknown parameters $a_{1}$and
$a_{2}$ provided that the parameter $\sigma$ is known from the large-order
behavior. Then we use the Meijer G-function representation  given by:%
\begin{equation}
_{\text{ }2}F_{0}\left(  {a_{1},a_{2};\ ;-\sigma x}\right)  =\frac{1}%
{\Gamma\left(  a_{1}\right)  \Gamma\left(  a_{2}\right)  }\MeijerG*{1}{2}{2}%
{1}{1-a_{1},1-a_{2}}{0}{-\sigma x,}%
\end{equation}
to obtain an approximant for the quantity $Q(x)$ in Eq.(\ref{pertQ}) for $M=2$. 

For the $M=3$ approximant ($_{3}F_{1}(a_{1},a_{2},a_{3})
;b_{1}\ ;-\sigma x)$, we have the equations:

\begin{align}
-\frac{a_{1}a_{2}a_{3}}{b_{1}}\sigma &  =k_{1},\nonumber\\
\frac{1}{2}\frac{a_{1}\left(  1+a_{1}\right)  a_{2}\left(  1+a_{2}\right)
a_{3}\left(  1+a_{3}\right)  }{b_{1}\left(  1+b_{1}\right)  }\left(
-\sigma\right)  ^{2}  &  =k_{2},\nonumber\\
\frac{1}{6}\frac{a_{1}\left(  1+a_{1}\right)  \left(  2+a_{1}\right)
a_{2}\left(  1+a_{2}\right)  \left(  2+a_{2}\right)  a_{3}\left(
1+a_{3}\right)  \left(  2+a_{3}\right)  }{b_{1}\left(  1+b_{1}\right)  \left(
2+b_{1}\right)  }\left(  -\sigma\right)  ^{3}  &  =k_{3},\\
a_{1}+a_{2}+a_{3}-b_{1}-2  &  =b,\nonumber
\end{align}
to be solved for the four unknowns $a_{1},a_{2},a_{3}$ and $b_{1}.$
Thus we get the approximation of $Q(x)$ as:%
\begin{equation}
Q_3(x)\approx\frac{\Gamma(b_{1})}{\Gamma\left(  a_{1}\right)  \Gamma\left(
a_{2}\right)  \Gamma\left(  a_{3}\right)  }\MeijerG*{1}{3}{3}{2}{1-a_{1}%
,1-a_{2},1-a_{3}}{0,1-b_{1}}{-\sigma x}.
\end{equation}
Since   these approximants are representing   approximations to the quantity $Q(x)$ thus they are always  uncertain unless we reached the exact resummation result for which $Q_l=Q_{l+1}= \dots$. Since this is not the case in most of the problems for a give $M$ order of the given perturbation series,   one needs then to set a criteria for the error calculation which we will pursue in the next subsection.

\subsection{Error Calculations}\label{algo2}
In literature, there are two main general approaches for the error prediction. The principle of fastest
apparent convergence (PFAC), where one adopts the arbitrary resummation parameters to minimize the difference $|Q_l-Q_{l-1}|$. Here $Q_l$ is the $l^{th}$ resummation order of the quantity $Q$. Another approach, is the principle of minimal sensitivity (PMS) for which one selects a parameter $\alpha$ that makes the quantity $Q(\alpha)$ less sensitive or stationary. A condition to approach this is to find the optimal value of the arbitrary parameter $\alpha$ from the relation  $\frac{\partial Q(\alpha)}{\partial\alpha}=0$ \cite{error,error2,error3,ON17}. In fact, the resummation procedure we described above seems to include no arbitrary parameters as we have always a sufficient number of equations to solve for all the parameters. However, we can have more than one approximant to approximate the same quantity. For instance, the $7^{th}$ order  of the perturbation series of $Q$ can be approximated by the approximants:
 
 \begin{enumerate}[(i)]
	 \item $Q(x)\approx {}_{5}F_{3}(a_{1},a_{2},.....,a_{5}; b_1,b_2,b_3;x)$
	\item $Q(x)\approx 1+ c_1 x \left({}_{4}F_{2}(a_{1},a_{2},.....,a_{4}; b_1,b_2;x)\right)$
	\item $Q(x)\approx 1+ c_1 x+c_2 x^2+c_3 x^3\left( {}_{3}F_{1}(a_{1},a_{2},a_{3}; b_1;x)\right)$
	\item $Q(x)\approx 1+ c_1 x+c_2 x^2+c_3 x^3+c_4 x^4+c_5 x^5\left({}_{2}F_{0}(a_{1},a_{2},; ;x)\right).$
 \end{enumerate}

	Likewise, the sixth order can be approximated by three different approximants:
  \begin{enumerate}[(a)]
	 	\item $Q(x)\approx  {}_{4}F_{2}(a_{1},a_{2},.....,a_{4}; b_1,b_2;x)$
	\item $Q(x)\approx 1+ c_1 x+c_2 x^2 \left( {}_{3}F_{1}(a_{1},a_{2},a_{3}; b_1;x)\right)$
	\item $Q(x)\approx 1+ c_1 x+c_2 x^2+c_3 x^3+c_4 x^4 \left({}_{2}F_{0}(a_{1},a_{2},; ;x)\right).$
 \end{enumerate}
 All these approximants are legal and can all be parametrized to give the needed features of the given perturbation series. One  can then vary the arbitrary   parameters $k$ in the seventh order approximant ${}_{k}F_{k-2}$ and $l$ in the sixth order approximant ${}_{l}F_{l-2}$ and selects the pair $(k,l)$ that minimizes the quantity $|Q_k(x)-Q_l(x)|$. In other words the pair $(k,l)$ are chosen to satisfy PFAC. The PFAC as well as PMS can be merged to determine the error in the resummation result \cite{ON17,error}. To do that for our resummation, we have to introduce arbitrary parameters into the approximants. For instance,  one can use the  approximant  $   {}_{5}F_{3}(a_{1},a_{2},.....,a_{5}; b_1,b_2,b_3;x)$ also to approximate the eighth order series. However, in all cases under investigation we have seven orders only and thus one can have any of the eight parameters as an arbitrary one determined by PMS. We select one of the numerator parameters $a_i$ as an arbitrary one for which one optimizes the quantity $Q(x,a_i)$ for its variation. The point is that the algorithm can give good approximation for the strong-coupling parameters ($-a_i$) \cite{Abo-large} and thus we know them approximately from the seventh order. So we can vary any of $a_i$ about their approximate values obtained from the known seven-loop approximant. In Ref.\cite{error}, the following error formula:
\begin{equation}
\Delta Q=|Q^{opt}_l-Q^{opt}_{l-1}|,\label{errort1}
\end{equation} 
has been used for   error calculation and $Q^{opt}_l$ is taken as the approximation for the quantity $Q$. Assume now we use the approximant $Q_5(x)\approx  {}_{4}F_{2}(a_{1},a_{2},.....,a_{4}; b_1,b_2;x)$ to approximate the fifth order of the perturbation series.  Assume also that we take $a_{1}$ as an arbitrary parameter. In fact one can do that and find $Q^{opt}_5(x)$ and also can find $Q^{opt}_6(x)$ similarly and calculate the error from the above formula. However, all the parameters in  ${}_{4}F_{2}(a_{1},a_{2},.....,a_{4}; b_1,b_2;x)$ can also be found from the known sixth order. So instead of taking  $Q^{opt}_5(x)$ we can replace it by $Q_6(x)$. Accordingly, we will variate only for the last order   approximant   which can be considered as an approximation for the unknown seventh order (say). Consequently, we shall  apply a modified form for the error as:  
\begin{equation}
\Delta Q=\frac{|Q_6-Q5|)+|Q_{opt}-Q_6|}{2} ,\label{errort}
\end{equation} 
We will take as our approximate quantity for $Q$ the quantity $(Q_6+Q_{opt})/2$ assuming we have only known six orders from the perturbation series.
 The formula we set for the error merges the PFAC and PMS in a fair way as it links the last three orders ($Q^{opt}_4$ is  replaced by $Q_5$, $Q^{opt}_5$ replaced by $Q_6$ as explained above).

\begin{figure}[t]
\begin{center}
\epsfig{file=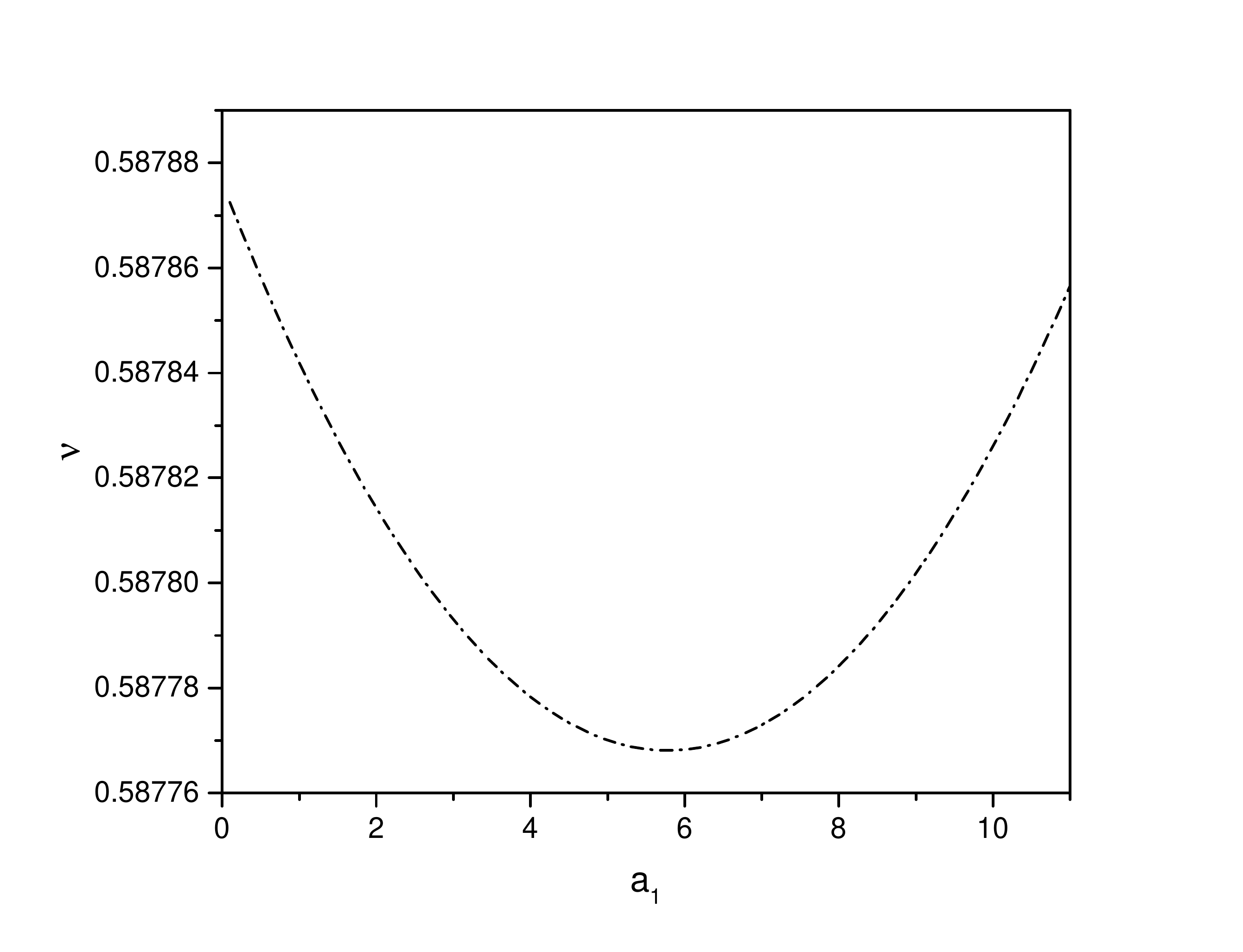,width=0.65\textwidth}
\end{center}
\caption{\textit{In this figure, we plot     the three dimensional exponent $\nu$ of the self-avoiding walks model ($N=0$) approximated by the  eight parameters approximant $\left(2\ {}_{5}F_{3}(a_{1},a_{2},.....,a_{5}; b_1,b_2,b_3;-\frac{3}{8}\varepsilon\right)$  versus $a_1$ as an arbitrary parameter.}} 
\label{opt} 
\end{figure}

Let us detail the algorithm by considering  a specific example. The seven-loop perturbation series of the reciprocal of the critical exponent $\nu$ for  the  self-avoiding walks $(N=0)$ is given by Eq.(\ref{nueps0}) (below). We found that the pair of approximants that best represents the sixth and seven loops is:  
\begin{align}
  \nu_7^{-1}   &  \approx 2 -\frac{1}{4} \varepsilon -\frac{11}{128} \varepsilon^2+0.114425 \varepsilon^3{\ }_{3}F_{1}(a_{1},a_{2},,a_{3}; b_1 ;-\sigma \varepsilon),\nonumber\\
  \nu_6^{-1}   &  \approx 2  {\ }_{4}F_{2}(a_{1},a_{2},.....,a_{4}; b_1,b_2 ;-\sigma \varepsilon),\label{sutablen0}
\end{align}        
where they give the results $\nu_7=0.587633, \nu_6=0.587439$ for $\varepsilon=1$ (three dimensions). For the optimization process we use an approximant that has the maximum available number of parameters with only one numerator parameter taken   arbitrary. In this case, it is the approximant $2 \ {}_{5}F_{3}(a_{1},a_{2},.....,a_{5}; b_1,b_2,b_3;-\sigma \varepsilon)$ where in figure \ref{opt} we plot the variation of this approximant  versus one of the numerator parameters while the other seven parameters are determined from the seven coefficients in Eq.(\ref{nueps0}). The value of $\nu$ at the optimized value of the parameter (minimum in this figure)     is considered as an approximation for the prediction from the unknown eighth order. The  parameter values are obtained as $a^{opt}_1=6.00000, a_2=-1.18738, a_3=-3.84527 , a_4= -0.015973, a_5=6.81113 , b_1= 1.0881-1.06875i, b_2=b^*_1,b_3=-3.84360$ and  the optimized $\nu$ exponent is then obtained from:

\begin{equation}
  \nu_{opt}=2\frac{\prod_{i=1}^{3}\Gamma\left(  b_{i}\right)  }{\prod
_{j=1}^{5}\Gamma\left(  a_{j}\right)  }\MeijerG*{1}{5}{5}{4}{1-a_{1}%
,\dots,1-a_{5}}{0,1-b_{1},1-b_{2},1-b_{3}}{-\sigma\varepsilon}.
\end{equation}
 Our prediction for $\nu_{opt}$ is $0.587773$  while the error is calculated from the relation:  
   \begin{equation}
\Delta \nu=(|\nu_7-\nu_6|)+|\nu_{opt}-\nu_7|)/2=0.000167.
\end{equation}
Accordingly the predicted value is $\nu=0.58770(17)$. This is a very precise result when compared to the conformal bootstrap result  $\nu=0.5877(12)$ in
Ref.\cite{BstrabN0}. In sec.\ref{resum7}, we will present the results for the exponents $\nu,\eta, \omega$ for the cases $N=0,1,2,3,4$.

A more crucial test for our algorithm can be offered by considering the two-dimensional ($\varepsilon=2$) case where exact results are well known. For this case we obtained the result $\nu=0.7512(39)$ which is compatible with  the exact result $\nu=0.75$ \cite{exact}. Taking into account that  the two-dimensional case is less divergent than the three dimensional one, our result shows that our    resummation results are very competitive. More two-dimensional exponents are presented in sec.\ref{2dim}. 

We will apply the above mentioned algorithm  to resum the $\varepsilon^7$ series for the critical exponents of the $O(N)$-symmetric model. Up to the best of our knowledge,   the $\varepsilon$-expansion for these exponents (for all  $N$ cases studied here and for the same exponents)  is not available so far in literature. So in the following section, we shall extract them first from the recent seven-loop calculations in Ref.\cite{7L}.
  
\section{\boldmath{\large{$\varepsilon$}}-Expansion for the seven-loop critical exponents of the
O(N)-symmetric Model}\label{7L-eps}
 For the $O(N)$-vector model , the Lagrangian density is given by:%
\begin{equation}
\mathcal{L=}\frac{1}{2}\left(  \partial\Phi\right)  ^{2}+\frac{m^{2}}{2}%
\Phi^{2}+\frac{16\pi^2 g}{4!}\Phi^{4},
\end{equation}
where $\Phi=\left(  \phi_{1},\phi_{2},\phi_{3},...........\phi_{N}\right)  $
is an N-component field. This Lagrangian    obeys an $O(N)$ symmetry where $\Phi
^{4}=\left(  \phi_{1}^{2}+\phi_{2}^{2}+\phi_{3}^{2}+...........\phi_{N}%
^{2}\right)  ^{2}$. In $4-\varepsilon$ dimensions within the minimal
subtraction technique, Oliver Schnetz has obtained the seven-loops order ($g$-expansion) for
the renormalization group functions $\beta,$ $\gamma_{m^{2}}$ and
$\gamma_{\phi}$ \cite{7L}. Here $\beta$ is the famous $\beta$-function that
determines the flow of the coupling in terms of mass scale, $\gamma_{m^{2}}$ is
the mass anomalous dimension and $\gamma_{\phi}$ represents the field anomalous dimension. In the following subsections, we list the corresponding seven-loop $\varepsilon$-expansion for each individual exponent for the cases $N=0,1,2,3,4$, respectively.     

\subsection{The seven-loop \boldmath{ $\varepsilon$}-expansion for self-avoiding walks $(N=0)$}

For $N=0$, we have the results \cite{7L}:
\begin{align}
\beta\left(  g\right)   &  \approx-\varepsilon g+2.667g^{2}-4.667g^{3}%
+25.46g^{4}-200.9g^{5}+2004g^{6}-23315g^{7}+303869g^{8},\nonumber\\
\gamma_{\phi}\left(  g\right)   &  \approx0.05556g^{2}-0.03704g^{3}%
+0.1929g^{4}-1.006g^{5}+7.095g^{6}--57.74g^{7},\label{N0-g}\\
\gamma_{m^{2}}\left(  g\right)   &  \approx-0.6667g+0.5556g^{2}-2.056g^{3}%
+10.76g^{4}-75.70g^{5}+636.7g^{6}-6080g^{7}.\nonumber
\end{align}
 The recipe to extract the corresponding $\varepsilon$-expansion  is direct where we solve the
equation $\beta\left(  g\right)  =0$ (fixed point) for the critical coupling
$g_{c}$ as a function of $\varepsilon$ and then substitute in the equations for
$\gamma_{\phi}\left(  g_{c}\right)  $ and $\gamma_{m^{2}}\left(  g_{C}\right)
$. Note that the critical exponents $\nu$ and $\eta$ are obtained from the relations
$\nu=\left[  2+\gamma_{m^{2}}\left(  g_{c}\left(  \varepsilon\right)
\right)  \right]  ^{-1}$ and $\eta\left(  \varepsilon\right)  =2\gamma_{\phi
}\left(  g_{c}\left(  \varepsilon\right)  \right)  $ while the correction to
scaling exponent $\omega$ is given as $\omega=\beta^{^{\prime}}\left(
g_c\right)  $. In Ref.\cite{epsilon7}, the method of  Lagrange inversion has been used to get
the exact seven-loop $\varepsilon$-expansion coefficients and has been applied to
the $N=1$ case but there the series has been obtained  for $\nu$ while here we list
the series for $\nu^{-1}$. However, here we will obtain the $\varepsilon
-$series by solving the equation $\beta\left(  g\right)  =0$ implicitly and
then expand the implicit solution as a power series in $\varepsilon$ keeping
only orders up to $O(\varepsilon^7)$. As we will see, our results are compatible  with those
obtained in Ref.\cite{epsilon7} for $\eta$ and $\omega$ for $N=1$. For
$\nu^{-1}, \eta$ and $\omega$ for $N=0,1,2,3$ and $4$, we found that our
results are compatible with the five-loop results available  in Ref.\cite{Kleinert-Borel} and
six-loop series (after proper scaling) in Ref.\cite{ON17}.   

For $N=0$ and after solving the equation $\beta(g(\varepsilon))$=0, we get the result:
\begin{equation}
g_{c}=0.37500\,\varepsilon+0.24609\varepsilon^{2}-0.18043\varepsilon^{3}+0.36808\allowbreak\varepsilon
^{4}-1.\,\allowbreak2576\varepsilon^{5}+5.\allowbreak0625\varepsilon
^{6}-23.\,\allowbreak392\varepsilon^{7},
\label{gc-N0}%
\end{equation}
and%

\begin{equation}
\nu^{-1}=2.0000-0.25000\varepsilon-0.08594\varepsilon^{2}+0.11443\varepsilon
^{3}-0.28751\varepsilon^{4}+0.95613\varepsilon^{5}-3.8558\varepsilon
^{6}+17.784\varepsilon^{7}, \label{nueps0}%
\end{equation}

\begin{equation}
\eta=0.015625\varepsilon^{2}+0.016602\varepsilon^{3}-0.0083675\varepsilon
^{4}+0.026505\varepsilon^{5}-0.090730\varepsilon^{6}+0.37851\varepsilon^{7},
\label{etaeps0}%
\end{equation}

\begin{equation}
\omega=1.0000\varepsilon-0.65625\varepsilon^{2}+1.8236\varepsilon^{3}-6.2854\varepsilon
^{4}+26.873\varepsilon^{5}-130.01\varepsilon^{6}+692.10\varepsilon^{7}.
\label{omegaeps0}%
\end{equation}
\subsection{ The \boldmath{ $\varepsilon$}-expansion for Ising-like model$(N=1)$}

In this case, the seven-loop $\beta-$function is presented in Ref.\cite{7L} as:
\begin{equation}
\beta\left(  g\right)  \approx-\varepsilon g+3.000g^{2}-5.667g^{3}%
+32.55g^{4}-271.6g^{5}+2849g^{6}-34776g^{7}+474651g^{8},
\end{equation}
and
\begin{equation}
\gamma_{m^{2}}\left(  g\right)  \approx-g+0.8333g^{2}-3.500g^{3}%
+19.96g^{4}-150.8g^{5}+1355g^{6}-13760g^{7},
\end{equation}
while
\begin{equation}
\gamma_{\phi}\left(  g\right)  \approx0.08333g^{2}-0.06250g^{3}+0.3385g^{4}%
-1.926g^{5}+14.38g^{6}-124.2g^{7}.
\end{equation}
Solving the equation $\beta\left(  g\right)  =0$,   we get the critical
coupling as:
\[
g_{c}=0.33333\varepsilon+0.20988\varepsilon^{2}-0.13756\varepsilon
^{3}+0.26865\varepsilon^{4}-0.84368\varepsilon^{5}+3.1544\varepsilon
^{6}-13.483\varepsilon^{7}.
\]
Substituting this form in $\gamma_{m^{2}}\left(  g_{c}\right)  $ and keep
orders up to $O(\varepsilon^{7})$ only we get:
\begin{align}
\nu^{-1}  &  =2.0000-0.33333\varepsilon-0.11728\varepsilon^{2}%
+0.12453\varepsilon^{3}-0.30685\varepsilon^{4}\label{nueps1}\\
&  +0.95124\varepsilon^{5}-3.5726\varepsilon^{6}+15.287\varepsilon
^{7}.\nonumber
\end{align}
 Similarly, the forms for $\eta$ and $\omega$ can be obtained as:%
\begin{equation}
\eta=0.018519\varepsilon^{2}+0.018690\varepsilon^{3}-0.0083288\varepsilon
^{4}+0.025656\varepsilon^{5}-0.081273\varepsilon^{6}+0.31475\varepsilon^{6},
\label{etaeps1}%
\end{equation}

\begin{equation}
\omega=\varepsilon-0.62963\varepsilon^{2}+1.61822\varepsilon^{3}%
-5.23514\varepsilon^{4}+20.7498\varepsilon^{5}-93.1113\varepsilon
^{6}+458.742\varepsilon^{7}. \label{omegaeps1}%
\end{equation}
\subsection{The  \boldmath{ $\varepsilon$}-expansion for $N=2$ ($XY$ universality
class)}

For $N=2$, the renormalization group functions are obtained in Ref.\cite{7L} as :%
\begin{equation}
\beta\approx-\varepsilon g+3.333g^{2}-6.667g^{3}+39.95g^{4}-350.5g^{5}%
+3845g^{6}-48999g^{7}+696998g^{8},
\end{equation}
\begin{equation}
\gamma_{m^{2}}\approx-1.333g+1.111g^{2}-5.222g^{3}+31.87g^{4}-255.8g^{5}%
+2434g^{6}-26086g^{7},
\end{equation}%
\begin{equation}
\gamma_{\phi}\approx0.11111g^{2}-0.09259g^{3}+0.5093g^{4}-3.148g^{5}%
+24.71g^{6}-224.6g^{7}.
\end{equation}
We extracted from these equations the following forms for $g_{c}$ and the
exponents $\nu,\eta$ and $\omega:$%
\begin{equation}
g_{c}=0.30000\varepsilon+0.18000\varepsilon^{2}-0.10758\varepsilon
^{3}+0.20502\varepsilon^{4}-0.59124\varepsilon^{5}+2.0719\varepsilon
^{6}-8.2614\varepsilon^{7},
\end{equation}
\begin{align}
\nu^{-1}  &  =2.0000-0.40000\varepsilon-0.14000\varepsilon^{2}%
+0.12244\varepsilon^{3}-0.30473\varepsilon^{4}\label{nueps2}\\
&  +0.87924\varepsilon^{5}-3.1030\varepsilon^{6}+12.419\varepsilon
^{7},\nonumber
\end{align}
\begin{equation}
\eta=0.020000\varepsilon^{2}+0.019000\varepsilon^{3}-0.0078936\varepsilon
^{4}+0.023209\varepsilon^{5}-0.068627\varepsilon^{6}+0.24861\varepsilon^{7},
\label{etaeps2}
\end{equation}
and
\begin{equation}
\omega=\varepsilon-0.60000\varepsilon^{2}+1.4372\varepsilon^{3}%
-4.4203\varepsilon^{4}+16.374\varepsilon^{5}-68.777\varepsilon^{6}%
+316.48\varepsilon^{7}. \label{omegaeps2}%
\end{equation}
\subsection{The \boldmath{ $\varepsilon$}-expansion for the exponents $\nu$,\ $\omega$\  and
$\eta$ in the Heisenberg universality class $(N=3)$}

The renormalization group functions can be generated using the maple package
in Ref.\cite{7L} as:%
\begin{equation}
\beta\left(  g\right)  \approx-\varepsilon g+3.667g^{2}-7.667g^{3}%
+47.65g^{4}-437.6g^{5}+4999g^{6}-66243g^{7}+978330g^{8},
\end{equation}
\begin{equation}
\gamma_{\phi}\approx0.1389g^{2}-0.1273g^{3}+0.6993g^{4}-4.689g^{5}%
+38.44g^{6}-365.9g^{7},
\end{equation}
and
\begin{equation}
\gamma_{m^{2}}\left(  g\right)  \approx-1.667g+1.389g^{2}-7.222g^{3}%
+46.64g^{4}-394.9g^{5}+3950^{6}-44412g^{7}.
\end{equation}
 From these functions, one can obtain the following forms for the critical
quantities $g_{c},\nu,\eta$ and $\omega$ as:
\begin{equation}
g_{c}=0.27273\varepsilon+0.15552\varepsilon^{2}-0.086255\varepsilon
^{3}+0.16154\varepsilon^{4}-0.42963\varepsilon^{5}+1.4210\varepsilon
^{6}-5.3221\varepsilon^{7},
\end{equation}
\begin{align}
\nu^{-1}  &  =2.0000-0.45455\varepsilon-0.15590\varepsilon^{2}%
+0.11507\varepsilon^{3}-0.29360\varepsilon^{4}\label{nueps3}\\
&  +0.78994\varepsilon^{5}-2.6392\varepsilon^{6}+9.9452\varepsilon
^{7},\nonumber
\end{align}
\begin{equation}
\eta=0.020661\varepsilon^{2}+0.018399\varepsilon^{3}-0.0074495\varepsilon
^{4}+0.020383\varepsilon^{5}-0.057024\varepsilon^{6}+0.19422\varepsilon^{7},
\label{etaeps3}%
\end{equation}
and
\begin{equation}
\omega=1.0000\varepsilon-0.57025\varepsilon^{2}+1.2829\varepsilon
^{3}-3.7811\varepsilon^{4}+13.182\varepsilon^{5}-52.204\varepsilon
^{6}+226.02\varepsilon^{7}. \label{omegaeps3}%
\end{equation}
\subsection{The seven-loop  \boldmath{ $\varepsilon$}-expansion for the $O(4)$-symmetric
case}

For $N=4$, we have the seven-loops $g$-series as :%
\begin{equation}
\beta\left(  g\right)  \approx-\varepsilon g+4.000g^{2}-8.667g^{3}%
+55.66g^{4}-533.0g^{5}+6318g^{6}-86768g^{7}+1.326\times10^{6}g^{8},
\end{equation}%
\begin{equation}
\gamma_{m^{2}}\left(  g\right)  \approx-2.000g+1.667g^{2}-9.500g^{3}%
+64.39g^{4}-571.9g^{5}+5983g^{6}-70240g^{7},
\end{equation}
and
\begin{equation}
\gamma_{\phi}\approx0.1667g^{2}-0.1667g^{3}+0.9028g^{4}-6.563g^{5}%
+55.93g^{6}-555.2g^{7}.
\end{equation}
Our prediction for the corresponding $\varepsilon-$expansion for the critical
coupling and exponents are of the form:
\[
g_{c}=0.25000\varepsilon+0.13542\varepsilon^{2}-0.070723\varepsilon
^{3}+0.13030\varepsilon^{4}-0.32185\varepsilon^{5}+1.0099\varepsilon
^{6}-3.5738\varepsilon^{7},
\]
\begin{align}
\nu^{-1}  &  =2.0000-0.5\varepsilon-0.166667\varepsilon^{2}%
+0.105856\varepsilon^{3}-0.278661\varepsilon^{4}\label{nueps4}\\
&  +0.702167\varepsilon^{5}-2.23369\varepsilon^{6}+7.97005\varepsilon
^{7},\nonumber
\end{align}%
\begin{equation}
\eta=0.020833\varepsilon^{2}+0.017361\varepsilon^{3}-0.0070852\varepsilon
^{4}+0.017631\varepsilon^{5}-0.047363\varepsilon^{6}+0.15219\varepsilon^{7},
\label{etaeps4}%
\end{equation}
and
\begin{equation}
\omega=\varepsilon-0.541667\varepsilon^{2}+1.15259\varepsilon^{3}%
-3.27193\varepsilon^{4}+10.8016\varepsilon^{5}-40.5665\varepsilon
^{6}+166.256\varepsilon^{7}. \label{omegaeps4}%
\end{equation}

 It is well known that the $\varepsilon-$series is divergent and has an
asymptotic large-order behavior of the type shown in Eq.(\ref{large-order})
where \cite{Kleinert-Borel}: $\sigma=-\frac{3}{N+8}$, $b_{\nu^{-1}}=4+\frac
{N}{2}$ , $b_{\eta}=3+\frac{N}{2}$ and $b_{\omega}=5+\frac{N}{2}.$
Accordingly, the suitable hypergeometric approximant is of the form
$_{p}F_{p-2}(a_{1},a_{2},....,a_{p};b_{1},b_{2},....b_{p-2};-\sigma x)$. In
the following section, we list the three dimensional ($\varepsilon$=1) resummation results for $N=0,1,2,3,4$ cases for
the exponents $\nu,\eta$ and $\omega$ within the $O(N)$-symmetric $\phi^4$ model.

\section{ hypergeometric-Meijer Resummation of the  \boldmath{\large{ $\varepsilon^7$}}
perturbation series}\label{resum7}

In this section, we present the resummation results for the $\varepsilon^7$-series but numerical values in this section are for $\varepsilon=1$. For the series representing $\nu^{-1}$ in Eqs.(\ref{nueps0},\ref{nueps1}%
,\ref{nueps2},\ref{nueps3},\ref{nueps4}), the suitable approximants are chosen to minimize the difference $\Delta \nu=|\nu_{7,k}-\nu_{6,\grave{k}}|$ where $k,\acute{k}$ are integers characterizing the hypergeometric approximant used. Then, one optimizes  for one of the numerator parameters as we explained in sec.\ref{algo}. For optimization, we select the approximant:
\begin{equation}
 \nu^{-1}_{opt}= 2\ {}_{5}F_{3}\left(a^{opt}_{1},a_{2},.....,a_{5}; b_1,b_2,b_3;-\sigma\varepsilon\right)
\end{equation}
as the one that   maximizes the number of parameters with only one of them to be taken arbitrary. Our prediction for the critical exponent $\nu$ is taken as $\nu=(\nu_7+\nu_{opt})/2$ while the error is calculated using the formula  $(|\nu_{7,k}-\nu_{6,\acute{k}}|+\nu_{opt}-\nu_{7,k}|)/2$. The recipe is also taking into account the analytic continuation of the divergent hypergeometric series $\ _{k}F_{k-2}(a_{1},...a_{k};b_{1},b_{2},....b_{k};-\sigma\varepsilon)$ as:     
\begin{equation}
  _{k}F_{k-2}\left(a_{1},...a_{k};b_{1},b_{2},....b_{k-2};-\sigma\varepsilon\right)=\frac{\prod_{i=1}^{k-2}\Gamma\left(  b_{i}\right)  }{\prod
_{j=1}^{k}\Gamma\left(  a_{j}\right)  }\MeijerG*{1}{k}{k}{k-1}{1-a_{1}%
,\dots,1-a_{k}}{0,1-b_{1},1-b_{2},\dots1-b_{k-2}}{-\sigma\varepsilon}.
\end{equation}
The same recipe will be followed for    the series representing the critical exponent $\omega$ in
Eqs.(\ref{omegaeps0},\ref{omegaeps1},\ref{omegaeps2},\ref{omegaeps3},\ref{omegaeps4}) except that the approximant taken for optimization is 
\begin{equation}
 \omega_{opt}=\ {}_{5}F_{3}\left(a^{opt}_{1},a_{2},.....,a_{5}; b_1,b_2,b_3;-\sigma\varepsilon\right)-1.
\end{equation}

For the   series given for the exponent $\eta$ in Eqs.(\ref{etaeps0},\ref{etaeps1}%
,\ref{etaeps2},\ref{etaeps3},\ref{etaeps4}), we also follow the same recipe but with optimizing the approximant:
\begin{equation}
\eta_{opt}={}_{5}F_{3}(a_{1},a_{2},.....,a_{5}; b_1,b_2,b_3;-\sigma \varepsilon)-\left(1-\sigma\varepsilon\frac{\prod_{j=1}^{5}a_{j}}{\prod_{i=1}^{3}b_{i}}\right).
\end{equation}
In the following, a subsection will be dedicated to the resummation of the exponents for every    $N$ . 
\subsection{ Seven-loop (\textbf{\large{$\varepsilon^7$}}) Resummation results  for self-avoiding walks $(N=0)$}

The seven-loop series for the reciprocal of the critical exponent $\nu$   is represented by Eq.(\ref{nueps0}). As detailed in sec.\ref{algo}, we found the suitable pair of approximants for the six and seven loops as in Eq.(\ref{sutablen0}) from which we obtained the results $\nu_7=0.587633, \nu_6=0.587439$ while our optimized result is $\nu_{opt}=0.587773$. These results lead to our prediction  as $\nu=0.58770(17)$ which is compatible with the recent conformal bootstrap result $\nu=0.5877(12)$ in Ref.\cite{BstrabN0} and also with the most precise  result  $\nu=0.5875970(4)$ from Monte Carlo simulations  in Ref.\cite{nuN0}.

For the critical exponent $\omega$ represented by the series in
Eq.(\ref{omegaeps0}),  we found that the best pair of  approximants that represent the six as well as the seven loops is:
\begin{align}
  \omega_7 &=   \varepsilon - 0.65625 \varepsilon^2 + 1.82361 \varepsilon^3{\ }_{3}F_{1}(a_{1},a_{2},a_{3}; b_1 ;-\sigma \varepsilon),\nonumber\\
  \omega_6   &=  \varepsilon - 0.65625 \varepsilon^2 {\ }_{3}F_{1}(a_{1},a_{2},a_{3}; b_1 ;-\sigma \varepsilon)
	\label{sutomega0}.
\end{align}
This pair has been chosen to minimize the difference $\Delta \omega_{6,7}=|\omega_7-\omega_6|$. Note that the parameters in each approximants are not the same but they take their own values based on matching expansions up to each order.
This gives the results $\omega_6=0.846452,\omega_7=0.846974$ while the optimized exponent is given by $\omega_{opt}=0.84990$. Accordingly, our prediction is 
$\omega=0.8484(17)$. In fact, the calculations from  Monte Carlo simulations in Ref.\cite{nuN0} gives the result $\omega=\frac{\Delta_{1}}{\nu}=0.899(12)$ while the six-loop Borel  with conformal mapping resummation \cite{ON17} turns the result $\omega=0.841(13).$

For the critical exponent $\eta$, it has the seven-loop perturbation series in
Eq.(\ref{nueps0}) and the suitable approximants are:

\begin{align}
  \eta_7 &=   \varepsilon\left( {\ }_{4}F_{2}(a_{1},a_{2},a_{3},a_{4}; b_1,b_2 ;-\sigma \varepsilon)-1\right),\nonumber\\
  \eta_6   &=  \frac{1}{64}\varepsilon^2 {\ }_{3}F_{1}(a_{1},a_{2},a_{3}; b_1 ;-\sigma \varepsilon),\label{suteta0}.
\end{align}
 These parametrization gives the result
$\eta_6=0.030336,\eta_7=0.030694$ while the optimized value is $\eta_{opt}=0.0317405$. Accordingly, our result is $\eta=0.03121(70)$. The bootstrap calculations gives the result $\eta=2\Delta_{\phi}-1=0.0282(4)$ \cite{BstrabN0} and the Monte Carlo result is $\eta=0.031043(3)$ \cite{nuN0E,ON17}. 

For the comparison with the predictions from  different other methods, our results for the three exponents 
are listed again in table \ref{7L0}.
\begin{table}[ht]
\caption{{\protect\scriptsize { The seven-loop ($\varepsilon$-expansion)   hypergeometric-Meijer ($\varepsilon^7$;HM)
resummation  results for the exponents $\nu,\eta$ and $\omega$  of the  self-avoiding walks model $(N=0)$.  The recent predictions of what is called self-consistent (SC) resummation algorithm introduced in Ref.\cite{Fractal} is also listed in the table. Besides,  we list results from conformal bootstrap (CB) calculations  \cite{BstrabN0}, Monte Carlo simulation (MC) for $\nu$ from Ref.\cite{nuN0E,ON17} and $\eta$ from Ref.\cite{nuN0}. Also the predictions of  the resummation of six-loop series using Borel with conformal mapping (BCM) algorithm ($\varepsilon^6$) from Ref.\cite{ON17} and five-loop ($\varepsilon^5$) from same reference is included.}}}%
\label{7L0}
\begin{tabular}{|l|l|l|l| }
\hline
\ \ Method\ \ & $\ \ \ \ \ \  \nu$ & $\ \ \ \ \ \ \eta$ & $\ \ \ \ \ \omega$   \\ \hline
\ \ \begin{tabular}[c]{@{}l@{}}$\varepsilon^7$; HM:  This work\\ \ \ SC\\ \ \ \ CB\\  \ \ \ MC\\ $\varepsilon^6$: BCM \\ $\varepsilon^5$: BCM \end{tabular}\ \ & \begin{tabular}[c]{@{}l@{}}0.58770(17)\\  0.5874(2) \\ 0.5877(12)\\  0.5875970(4)\\ 0.5874(3)\\0.5873(13)\end{tabular} & \begin{tabular}[c]{@{}l@{}}0.03121(70)\\0.0304(2)\\ 0.0282(4)\\  0.031043(3)\\ 0.0310(7)\\0.0314(11)\end{tabular} & \begin{tabular}[c]{@{}l@{}}0.8484(17)
 \\ 0.846(15)\\ \ \ \  \textbf{---}\\ 0.899(12)\\0.841(13)\\0.835(11)\end{tabular}  \\ \hline 

\end{tabular}%
\end{table}

\subsection{Seven-loop Resummation results for Ising-like universality class ( $N=1$)}

The perturbation series for critical exponent $\nu(\nu^{-1})$ of the Ising-like model up to 
 $\varepsilon^7$  is given by Eq.(\ref{nueps1}). The suitable approximants for six and seven loops are:
\begin{align}
  \nu_6 &=   2\ {\ }_{4}F_{2}(a_{1},a_{2},a_{3},a_{4}; b_1,b_{2} ;-\sigma \varepsilon),\nonumber\\
  \nu_7  &=  2\  {\ }_{5}F_{3}(a_{1},a_{2},\dots,a_{5}; b_1,b_{2},b_{3} ;-\sigma \varepsilon)\label{sutnu1}.
\end{align}
These approximants  give the results $\nu_6=0.629374,\nu_7=0.629732$ while $\nu_{opt}=0.629809$. Accordingly, our prediction is $\nu=0.62977(22)$.
  To get an idea about how accurate this result is, we list here results from recent non-perturbative methods like the Monte Carlo simulations 
which gives the  result  $\nu=0.63002(10)$ \cite{MC10}, the recent   non-perturbative renormalization group (NPRG) method \cite{NPRG} which turns the result    $\nu=0.63012(16)$ as well as the recent conformal bootstrap result  $\nu=0.62999(5)$ in Ref.\cite{Bstrab2}. In view of these non-perturbative calculations and in looking at  table \ref{7L1}, one can realize that the seven-loop resummation results has improved significantly  the six-loop results. In fact, this is a general trend in all of the seven-loop calculations in this work.

For the exponent $\eta$,  the suitable approximants are:
\begin{align}
  \eta_7 &=   0.0185185 \varepsilon^2+ 0.01869 \varepsilon^3 - 0.00832877\varepsilon^4+0.0207101\varepsilon^5\ {\ }_{2}F_{0}(a_{1},a_{2};  ;-\sigma \varepsilon),\nonumber\\
  \eta_6   &=  0.0185185 \varepsilon^2 {\ }_{3}F_{1}(a_{1},a_{2},a_{3}; b_1 ;-\sigma \varepsilon),\label{suteta1 }.
\end{align}
which give the result $\eta_6=0.0354512,\ \eta_7=0.0363063$. Also our optimized result is $\eta_{opt}=0.0367504$ and all of these results lead to the predicted value 
$\nu=0.03653(65)$ compared to Monte Carlo simulation result of $\eta=0.03627(10)$, NPRG result $\eta=0.0361(11)$ \cite {NPRG} and conformal bootstrap calculation of
$\eta=0.03631(3)$ \cite{Bstrab2} .

For the correction to scaling exponent $\omega$, the up to seven-loop order of
perturbation series is given by Eq.(\ref{omegaeps1}). The approximants used are
\begin{align}
  \omega_7 &=   {\ }_{5}F_{3}(a_{1},a_{2},\dots,a_{5}; b_1,b_{2},b_{3} ;-\sigma \varepsilon)-1,\nonumber\\
  \omega_6   &=  \varepsilon - 0.62963  \varepsilon^2 {\ }_{3}F_{1}(a_{1},a_{2},a_{3}; b_1 ;-\sigma \varepsilon),\label{sutomega1}
\end{align}
where we get the results $\omega_6=0.823572,\omega_7=0.823592$ while we get the optimized result $\omega_{opt}=0.822619$. Thus our prediction for the $\omega$ exponent is $\omega=0.82311(50)$. NPRG method gives the result  $\omega=0.832(14)$ \cite{NPRG} while  conformal
bootstrap has the result $\omega=0.8303(18)$ \cite{Bstrab2} and Monte Carlo
simulations predicts the value $\omega=0.832(6)$ \cite{MC10}.  Comparison with the predictions from more different  methods  is  listed in  table \ref{7L1}.

\begin{table}[ht]
\caption{{\protect\scriptsize { The seven-loop ($\varepsilon$-expansion)   hypergeometric-Meijer resummation  results ($\varepsilon^7$;HM) for the exponents $\nu,\eta$ and $\omega$  of the  Ising-like  model $(N=1)$.  The recent SC resummation results are listed for comparison.  Also we list conformal bootstrap calculations  from Ref.\cite{Bstrab2} and Monte Carlo simulation (MC)  from Ref.\cite{MC10}. In this table also, we list the six-loop ($\varepsilon^6$) resummation results from the Borel with conformal mapping (BCM)   from Ref.\cite{ON17} and five-loops ($\varepsilon^5$) from same reference. The recent results from the  non-perturbative renormalization group (NPRG) method \cite{NPRG} is listed last. }}}%
\label{7L1}
\begin{tabular}{|l|l|l|l| }
\hline
\ \ Method\ \ & $\ \ \ \ \ \  \nu$ & $\ \ \ \ \ \ \eta$ & $\ \ \ \ \ \omega$   \\ \hline
\ \ \begin{tabular}[c]{@{}l@{}}$\varepsilon^7$; HM:  This work\\  \ \ SC\\ \ \ \ CB\\  \ \ \ MC\\ $\varepsilon^6$: BCM \\ $\varepsilon^5$: BCM\\NPRG \end{tabular}\ \ & \begin{tabular}[c]{@{}l@{}}0.62977(22) \\0.6296(3) \\ 0.62999(5)\\  0.63002(10)\\ 0.6292(5)\\0.6290(20)\\ 0.63012(16)\end{tabular} & \begin{tabular}[c]{@{}l@{}} 0.03653(65)  \\0.0355(3) \\  0.03631(3)\\  0.03627(10)\\ 0.0362(6)\\0.0366(11)\\0.0361(11) \end{tabular} & \begin{tabular}[c]{@{}l@{}}0.82311(50)\\   0.827(13)\\ 0.8303(18)\\ 0.832(6)\\0.820(7)\\0.818(8)\\0.832(14)\end{tabular}  \\ \hline 

\end{tabular}%
\end{table}

\subsection{  Seven-loop Resummation results for $XY$ universality class ( $N=2$)}

For $N=2$, the perturbation series up to $\varepsilon^{7}$ for the critical
exponent $\nu$ is given by Eq.(\ref{nueps2}) while our suitable approximants are
\begin{align}
  \nu_6 &=  2-0.4\ {\ }_{3}F_{1}(a_{1},a_{2},a_{3}; b_1;-\sigma \varepsilon),\nonumber\\
  \nu_7  &=  2-0.4\  \varepsilon \ {}_{4}F_{2}(a_{1},...a_{4};b_{1},b_{2};-\sigma\varepsilon)\label{sutnu2}.
\end{align}
These approximants give the results $\nu_6=0.671347, \nu_7=0.670934$ while the optimized value is $\nu_{opt}=0.670582$. These results all predict the value $\nu=0.67076
(38)$ as our result for the $\nu$ exponent for $N=2$. Our prediction is compatible with the result from the microgravity experiment of  $\nu=0.6709(1)$  \cite{alphaxy,dispute1}. Also, it is very close (but not completely compatible) to the more precise Monte Carlo result is $\nu=0.67169(7)$ \cite{MC19} and the very recent bootstrap result $\nu=0.67175(10)$ \cite{dispute1,dispute}.  To compare with more other works, we mention that the  NPRG yields the prediction $\nu=0.6716(6)$ \cite{NPRG} while  the recent  conformal bootstrap prediction is $\nu=0.6719(11)$ \cite{Bstrab4}.

A note to be mentioned here is that our prediction of $\nu=0.67076(38)$ leads to the result $\alpha= -0.0123(11)$ where $\alpha$ is the critical exponent associated with the singularity in specific heat. The  zero gravity  liquid helium superfluid transition experiment gives the result -0.0127(3) \cite{alphaxy} which is compatible with our result. In taking into account that the six-loop Borel with conformal mapping  resummation result is $\alpha=-0.007(3)$, one can realize the significant improvement the resummation of the seven-loop adds. However, neither our result nor the experimental results are compatible with MC \cite{MC19} or CB \cite{dispute,dispute1} calculations.

For the critical exponent $\eta$, the seven-loop series is given in Eq.(\ref{etaeps2}). The suitable  hypergeometric-Meijer approximants are:
 \begin{align}
  \eta_7 &=   \varepsilon\left( {\ }_{4}F_{2}(a_{1},a_{2},a_{3},a_{4}; b_1,b_2 ;-\sigma \varepsilon)-1\right),\nonumber\\
  \eta_6   &= 0.02 \varepsilon^2 {\ }_{3}F_{1}(a_{1},a_{2},a_{3}; b_1 ;-\sigma \varepsilon).
	\label{suteta2}
\end{align}
These approximants give the results $\eta_6=0.0373266, \eta_7=	0.037753$ while $\eta_{opt}=0.038448$. Thus our prediction is $\eta=0.03810(56)$. The NPRG method predicted the value $\eta=0.0380(13)$ \cite{NPRG}, Monte Carlo simulations in Ref.\cite{MC19} gives the result $\eta=0.03810(8)$ and conformal bootstrap has the prediction $\eta=0.03852(64)$ \cite{Bstrab4}.

For the resummation of the critical exponent $\omega$ given by Eq.(\ref{omegaeps3}), we used the approximants:
\begin{align}
  \omega_7 &=   {\ }_{5}F_{3}(a_{1},\dots,a_{5}; b_1,b_2,b_3 ;-\sigma \varepsilon)-1,\nonumber\\
  \omega_6   &=  \varepsilon -0.6 \varepsilon^2 {\ }_{3}F_{1}(a_{1},a_{2},a_{3}; b_1 ;-\sigma \varepsilon),
	\label{sutomega2}
\end{align}
where they give the results $\omega_6=0.803962,\omega_7=0.801541$. Also, we get $\omega_{opt}=0.776887$. So our prediction is $\omega=0.789(13)$.  The result $\omega=0.789(4)$ has been shown using recent  Monte Carlo calculations \cite{MC19} and the prediction of conformal bootstrap calculations yields the result $\omega=0.811(10)$
\cite{Bstrab3,ON17} while the recent NPRG result is $\omega=0.791(8)$ \cite{NPRG}.

 In table \ref{7L2}, we list predictions from more different methods for the three exponents beside our predictions. 

\begin{table}[ht]
\caption{{\protect\scriptsize { The    hypergeometric-Meijer ($\varepsilon^7$;HM)
resummation  results for the  exponents $\nu,\eta$ and $\omega$   of the $O(2)$-symmetric model.  The recent SC results  are  taken from Ref.\cite{Fractal}. Other predictions are listed  from  conformal bootstrap calculations \cite{dispute,dispute1} for $\nu$  and   $\eta$ \cite{Bstrab4}, while $\omega$ result is taken from Refs.\cite{Bstrab3,ON17} and MC calculations  from Ref.\cite{MC19}. The six-loop BCM resummation ($\varepsilon^6$) from Refs.\cite{ON17} and the five-loops ($\varepsilon^5$) from same reference. In the last row we add the    NPRG results up to $O(\partial^4$) \cite{NPRG}.}}}%
\label{7L2}
\begin{tabular}{|l|l|l|l| }
\hline
\ \ Method\ \ & $\ \ \ \ \ \  \nu$ & $\ \ \ \ \ \ \eta$ & $\ \ \ \ \ \omega$   \\ \hline
\ \ \begin{tabular}[c]{@{}l@{}}$\varepsilon^7$; HM:  This work\\  \ \ \ SC\\ \ \ \ CB\\  \ \ \ MC\\ $\varepsilon^6$: BCM \\ $\varepsilon^5$: BCM\\NPRG \end{tabular}\ \ & \begin{tabular}[c]{@{}l@{}}0.67076(38) \\0.6706(2) \\ 0:67175(10)\\  0.67169(7)\\ 0.6690(10)\\0.6687(13)\\ 0.6716(6)\end{tabular} & \begin{tabular}[c]{@{}l@{}}0.03810(56)\\ 0.0374(3)\\  0.03852(64)\\  0.03810(8)\\ 0.0380(6)\\0.0384(10)\\0.0380(13) \end{tabular} & \begin{tabular}[c]{@{}l@{}}0.789(13)\\0.808(7) \\  0.811(10)\\ 0.789(4)\\0.804(3)\\0.803(6)\\0.791(8)\end{tabular}  \\ \hline 
\end{tabular}%
\end{table}

\subsection{Seven-loop Resummation results for Heisenberg universality class ( $N=3$)}

For the case of the  Heisenberg universality class ( $N=3$), the reciprocal of the critical exponent $\nu$ has the $\varepsilon^{7}$ perturbation series
given by Eq.(\ref{nueps3}). The approximants that minimize the difference $\Delta_{\nu}=|\nu_7-\nu_6|$ are found to be: 

\begin{align}
  \nu_6 &=  2 - 0.454545 \varepsilon - 0.155898 \varepsilon^2 \ {\ }_{3}F_{1}(a_{1},a_{2},a_{3}; b_1;-\sigma \varepsilon),\nonumber\\
  \nu_7  &=  2 - 0.454545 \varepsilon - 0.155898 \varepsilon^2+0.115071 \varepsilon^3 \ {\ }_{3}F_{1}(a_{1},a_{2},a_{3}; b_1;-\sigma \varepsilon)\label{sutnu3}.
\end{align}
 The parametrization of these approximants leads to the predictions $\nu_6=0.709151,\nu_7=	0.709212$ while our optimized result is  $\nu_{opt}=	0.708906$. These results all together give the prediction $\nu=0.70906(18)$. The conformal bootstrap calculations gives the value $\nu=0.7121(28)$ while the Monte Carlo simulations in Ref.\cite{MC11}  gives the result $\nu=0.7116(10)$ and the NPRG method has the value  $\nu=0.7114(9)$ \cite{NPRG}.

The series up to $\varepsilon^{7}$ for the exponent $\eta$ is given in Eq.(\ref{etaeps3}) where it has been approximated using:
  \begin{align}
  \eta_7 &=   \varepsilon\left( {\ }_{4}F_{2}(a_{1},a_{2},a_{3},a_{4}; b_1,b_2 ;-\sigma \varepsilon)-1\right),\nonumber\\
  \eta_6   &= \frac{5}{242} \varepsilon^2 {\ }_{3}F_{1}(a_{1},a_{2},a_{3}; b_1 ;-\sigma \varepsilon).
	\label{3}
\end{align}
 These approximations yield the prediction $\eta_6=0.0373004,\eta_7=	0.0376457$ while $\eta_{opt}$ gives the result	$0.0385369$. Accordingly, we have our prediction as $\eta=0.03809(62)$  compared to $\eta=0.0386(12)$ from bootstrap calculations in Ref. \cite{Bstrab4},  $\eta=0.0376(13)$ from NPRG in Ref.\cite{NPRG}  and $\eta=0.0378(3)$ predicted by Monte Carlo simulations in Ref.  \cite{MC11} .

The seven-loop $\varepsilon$-expansion for the exponent $\omega$ has been
obtained in the previous section in Eq.(\ref{omegaeps3}). This series has been
resummed through the use of the hypergeometric approximants:
\begin{align}
  \omega_7 &=   \varepsilon - 0.570248 \varepsilon^2 + 1.2829 \varepsilon^3{\ }_{3}F_{1}(a_{1},a_{2},a_{3}; b_1 ;-\sigma \varepsilon),\nonumber\\
  \omega_6   &=  \varepsilon  - 0.570248 \varepsilon^2 + 1.2829 \varepsilon^3 - 3.78111 \varepsilon^4  {\ }_{2}F_{0}(a_{1},a_{2} ;   ;-\sigma \varepsilon),
	\label{sutomega3}
\end{align}
where they give the results $\omega_6=0.761533,\omega_7=	0.775002$. Also, we obtained the result $\omega_{opt}=	0.752398$. Accordingly, our results for the coorection to scaling exponent is $\omega=0.764(18) $. For comparison, we list here the value $\omega=0.791(22)$ from  conformal bootstrap calculations  \cite{Bstrab3,ON17}, $\omega=0.773$ from  Monte Carlo  simulations  \cite{MC01} and $\omega=0.769(11)$ from NPRG method \cite{NPRG}. 

In table \ref{7L3}, we list more results from other methods to make it clear that the
hypergeometric-Meijer resummation algorithm though simple is  competitive
to other more sophisticated algorithms and methods.  

\begin{table}[ht]
\caption{{\protect\scriptsize { The seven-loop ($\varepsilon^7$)  hypergeometric-Meijer resummation  for the  exponents $\nu$ , $\eta$ and
 $\omega$  of the $O(3)$-symmetric model. Also we list the SC resummation results from Ref.\cite{Fractal}. The recent  results from  conformal bootstrap calculations are listed also where the values of  $\nu$  and $\eta$ are taken from Ref. \cite{Bstrab4} while $\omega$ from Refs.\cite{Bstrab3,ON17}.  For MC simulations  $\omega$ is taken from from Ref.\cite{MC01} while $\nu$ and $\eta$ are taken from from Ref.\cite{MC11}. The six-loop BCM resummation   is taken from Ref.\cite{ON17} and five-loops   from same reference. As in all of above tables, we list in the last row the very recent calculations from NPRG method \cite{NPRG} (up to $O(\partial^4$)). }}}%
\label{7L3}
\begin{tabular}{|l|l|l|l| }
\hline
\ \ Method\ \ & $\ \ \ \ \ \  \nu$ & $\ \ \ \ \ \ \eta$ & $\ \ \ \ \ \omega$   \\ \hline
\ \ \begin{tabular}[c]{@{}l@{}}$\varepsilon^7$; HM:  This work\\ \ \ \ SC\\ \ \ \ CB\\  \ \ \ MC\\ $\varepsilon^6$: BCM \\ $\varepsilon^5$: BCM\\NPRG \end{tabular}\ \ & \begin{tabular}[c]{@{}l@{}}0.70906(18)\\0.70944(2) \\ 0.7121(28)\\  0.7116(10)\\ 0.7059(20)\\0.7056(16)\\ 0.7114(9)\end{tabular} & \begin{tabular}[c]{@{}l@{}}0.03809(62)\\ 0.0373(3)\\0.0386(12)\\  0.0378(3)\\ 0.0378(5)\\0.0382(10)\\0.0376(13) \end{tabular} & \begin{tabular}[c]{@{}l@{}}0.764(18)\\ 0.794(4)\\0.791(22)\\ 0.773\\0.795(7)\\0.797(7)\\0.769(11)\end{tabular}  \\ \hline 

\end{tabular}%
\end{table}

\subsection{Seven-loop Resummation results for the   $N=4$ case }

Similar to the above cases, the seven-loop perturbation  series for the
exponent $\nu$ has been obtained in the previous section  in Eq.(\ref{nueps4}). The best choice for approximating the six and seven-loop series  is:
\begin{align}
  \nu_6 &=  2  -0.5 \varepsilon  - 0.166667 \varepsilon ^2 + 0.105856 \varepsilon ^3 - 0.278661 \varepsilon ^4 \ {\ }_{2}F_{0}(a_{1},a_{2} ;  ;-\sigma \varepsilon),\nonumber\\
  \nu_7  &=  2  -0.5 \varepsilon  - 0.166667 \varepsilon ^2 + 0.105856 \varepsilon ^3  \ {\ }_{3}F_{1}(a_{1},a_{2},a_{3}; b_1;-\sigma \varepsilon)\label{sutnu4}.
\end{align}
These approximations lead to the values $\nu_6=0.744328,\nu_7=	0.744488$.	 Besides, we get the optimized result as $\nu_{opt}=0.744017$. Accordingly, we have the result
 $\nu=0.74425 (32)$. The NPRG prediction is $\nu=0.7478(9)$ from Ref.\cite{NPRG}.  Also  the conformal bootstrap result  is $\nu=0.751(3)$ from Ref.\cite{Bstrab3} and   Monte Carlo simulations   gives the result $\nu=0.750(2)$ \cite{MC11}. 

For the critical exponent $\eta$ with perturbative result in Eq.(\ref{etaeps4}),
we used the approximants:\begin{align}
  \eta_7 &=   \varepsilon\left( {\ }_{4}F_{2}(a_{1},a_{2},a_{3},a_{4}; b_1,b_2 ;-\sigma \varepsilon)-1\right),\nonumber\\
  \eta_6   &= \frac{1}{48} \varepsilon^2 {\ }_{3}F_{1}(a_{1},a_{2},a_{3}; b_1 ;-\sigma \varepsilon).
	\label{4}
\end{align} So we have the results $\eta_6=0.0362091, \eta_7=	0.0364162, \eta_{opt}=	0.0369761$ which lead to our prediction   $\eta=0.03670(38)$. The NPRG result is $\eta=0.0360(12)$ \cite{NPRG} and Monte Carlo simulations for that case gives the result
$\eta=0.0365(3)$ \cite{MC11}  while recent bootstrap calculations gives the value $0.0378(32)$ \cite{Bstrab5}.

For the critical exponent $\omega$ represented by Eq.(\ref{omegaeps4}), we used the approximants: 

\begin{align}
  \omega_7 &=   {\ }_{5}F_{3}(a_{1},\dots,a_{5}; b_1,b_2,b_3 ;-\sigma \varepsilon)-1,\nonumber\\
  \omega_6   &=  \varepsilon  - 0.541667 \varepsilon^2 + 1.15259 \varepsilon^3 - 3.27193 \varepsilon^4 {\ }_{2}F_{0}(a_{1},a_{2} ; ;-\sigma \varepsilon),
	\label{sutomega4}
\end{align}
which yields the results $\omega_6=0.753489,\omega_7=0.75285$. Also, we have the optimized result as  	$\omega_{opt}=0.750977$. Accordingly, our results is 
$\omega=0.7519(13)$  compared to NPRG result of $\omega=0.761(12)$ \cite{NPRG} while the result of Monte Carlo simulation in Ref.\cite{MC01} is $\omega=0.765$   and $\omega=0.817(30)$ from   conformal bootstrap calculations  \cite{Bstrab3,ON17}.

\begin{table}[ht]
\caption{{\protect\scriptsize { The seven-loop ($\varepsilon^7$)  hypergeometric-Meijer resummation  for the  exponents $\nu$ , $\eta$ and
 $\omega$  of the $O(4)$-symmetric model compared to results from conformal bootstrap calculations \cite{Bstrab3,ON17} for $\nu$  and $\omega$  , while $\eta$ from Ref\cite{Bstrab5}, MC simulations  for  $\omega$ is taken from Ref.\cite{MC01} while $\nu$ and $\eta$  are from Ref.\cite{MC11}. Also, the six-loop BCM resummation ($\varepsilon^6$) is taken from Ref.\cite{ON17} and five-loops ($\varepsilon^5$) from same reference. NPRG results up to $O(\partial^4$) \cite{NPRG} are shown in the last row. }}}%
\label{7L4}
\begin{tabular}{|l|l|l|l| }
\hline
\ \ Method\ \ & $\ \ \ \ \ \  \nu$ & $\ \ \ \ \ \ \eta$ & $\ \ \ \ \ \omega$   \\ \hline
\ \ \begin{tabular}[c]{@{}l@{}}$\varepsilon^7$; HM:  This work\\   \ \ \ SC\\ \ \ \ CB\\  \ \ \ MC\\ $\varepsilon^6$: BCM \\ $\varepsilon^5$: BCM\\NPRG \end{tabular}\ \ & \begin{tabular}[c]{@{}l@{}}0.74425(32)\\ 0.7449(4)\\ 0.751(3)\\  0.750(2)\\ 0.7397(35)\\0.7389(24)\\ 0.7478(9)\end{tabular} & \begin{tabular}[c]{@{}l@{}}0.03670(38) \\0.0363(2)\\ 0.0378(32)\\  0.0360(3)\\ 0.0366(4)\\0.0370(9)\\0.0360(12) \end{tabular} & \begin{tabular}[c]{@{}l@{}} 0.7519(13)  \\ 0.7863(9)\\0.817(30)\\ 0.765 (30)\\0.794(9)\\0.795(6)\\0.761(12)\end{tabular}  \\ \hline 

\end{tabular}%
\end{table}

\section{Two-dimensional hypergeometric-Meijer resummation }\label{2dim}

In two dimensions or equivalently $\varepsilon=2$, there are two main
differences from the three dimensional case. The first is that for $N\geq2$,
there is no broken-symmetry phase \cite{plesito2}. For the other difference,
since $\varepsilon=2$ is a large value and the strong-coupling asymptotic
behavior of the $O(N)$ symmetric model is not known yet, one expects a
slower convergence of the resummation of the perturbation series. For the $g$ expansion, it has
been argued that the $\beta$ function  is not analytic at the fixed point
\cite{plesito2,nikel,sokolov} which in turn slows the convergence down too. The
effect of the non-analiticity of the $\beta$ function is higher in two
dimensions. This leads to inaccurate predictions for critical exponents from
the $g$ expansion in two dimensional case \cite{sokolov}. Accordingly, testing the
resummation algorithm for the  $\varepsilon=2$ case offers an interesting
point about the capability of the $\varepsilon$-expansion to predict reliable
results for that case. Apart from inaccurate resummation results from the $g$-expansion as well as previous results of the $\varepsilon$-expansion that needs more improvement, exact values for the two dimensional critical exponents are known and thus can be used to test the reliability of
any approximating  method.

For $N=0,$ our resummation result for the critical exponent $\nu$  is $0.751(4)$
which is compatible with  the exact result   assumed to be $0.75$ \cite{exact}. Note that the recent Borel
with conformal mapping resummation for six-loop yields the result $\nu=0.741(4)$ \cite{ON17}. 

For the critical exponent $\omega$, our prediction
is $1.96(46)$ while the exact value is $2$ \cite{exact,exact2} and the recent
six-loop resummation in Ref. \cite{ON17} gives the result $1.90(25)$. 

For $\eta$, we get the value $0.214(28)$ while the exact result is $(\frac
{5}{24})\approx0.20833...$ \cite{exact} and the six-loop resummation (BCM)  result   is $0.201(25)$ \cite{ON17}. One can realize that
our predictions show a clear improvement for the previous resummation results in
literature.

For Ising-like case ($N=1$), we obtained the result $\nu=0.976(13)$ compared to
the well known exact result $\nu=1$ \cite{Onsa} while BCM result for six loops
gives the value $\nu=0.952(14)$. For $\omega$ we get the result
$1.71(10)$ while the exact value is $\omega=1.75$ \cite{wn1} and BCM result is $1.71(9).$
Our prediction for $\eta$ is $0.243(25)$ while the exact value is $0.25$ \cite{exact} and
the six-loop BCM resummation result is $0.237(27)$.

\begin{table}[ht]
\caption{{\protect\scriptsize { The seven-loop ($\varepsilon^7$)  hypergeometric-Meijer resummation  for the   exponents $\nu$ , $\eta$ and
 $\omega$  for the Self-avoiding walks ($N=0$) and the Ising-like model ($N=1$) in two dimensions ($\varepsilon=2$). For comparison with other predictions, we list the results from  BCM  algorithm \cite{ON17} for six loops.  Exact results for $N=1$ for $\nu$ and $\eta$ are obtained in the seminal article in Ref.\cite {Onsa} and   $\omega$ from Ref.\cite{wn1}. For $N=0$, exact values for $\nu,\eta$ and $\omega$ are conjectured in Ref. \cite{exact}. }}}%
\label{7L01d2}
\begin{tabular}{|l|l|l|l|l|}
\hline
\ \ N \ \ & \ \ \ $\nu$ & \ \ \ $\eta$ & \ \ \ $\omega$ & Method \\ \hline
\ \ 0 &\begin{tabular}[c]{@{}l@{}} 0.751(4)\\ 0.741(4)\\0.75\end{tabular} &\begin{tabular}[c]{@{}l@{}} 0.214(28) \\0.201(25)\\0.20833...\end{tabular}  &\begin{tabular}[c]{@{}l@{}} 1.96(46) \\ 1.90(25)\\ \ \ \ 2 \end{tabular}& \ \ \begin{tabular}[c]{@{}l@{}}$\varepsilon^7$; HM:  This work \\ $\varepsilon^6$: BCM\\ Exact\end{tabular} \\ \hline
\ \ 1 &\begin{tabular}[c]{@{}l@{}} 0.976(13)\\0.952(14)\\ \ \ \ 1\end{tabular} &\begin{tabular}[c]{@{}l@{}} 0.243(25)\\0.237(27)\\ \ \ \ 0.25\end{tabular} & \begin{tabular}[c]{@{}l@{}}1.71(10) \\1.71(9)\\1.75\end{tabular} & \begin{tabular}[c]{@{}l@{}} $\varepsilon^7$; HM:  This work\\   $\varepsilon^6$: BCM\\ Exact \end{tabular}\\ \hline
\end{tabular}
\end{table}

\section{summary and conclusions}\label{conc}

The recent calculations of the seven-loop renormalization group functions of the $O(N)$-symmetric field theory ($g$-expansion) have motivated us to generate  the corresponding $\varepsilon$-expansion for the critical exponents $\nu,\eta$ and $\omega$.  Scaling relations can lead to other different critical exponents like the specific heat singularity exponent where it is given by $\alpha=3-D\nu$ with $D=4-\varepsilon$, the Euclidean space-time dimension. Getting the seven-loop order of the $\varepsilon$-expansion is important toward the improvement of the previous six-loop resummation results \cite{ON17,abo-expon}. In this work, we used our hypergeometric-Meijer algorithm \cite{abo-expon,Abo-large} to resum  the up to $\varepsilon^7$ series for the critical exponents from the $O(N)$-symmetric $\phi^4$ theory for $N=0,1,2,3,4$. The resummation results has shown clear improvement for the previous six-loop results. The most reflecting quantity for the improvement of the six-loop results is the specific heat critical exponent of the $XY$ model. Taking into account that the result $\alpha=-0.0127(3)$ from zero-gravity experiment in Ref.\cite{alphaxy}, the BCM six-loop result from ref.\cite{ON17} which is $\alpha=-0.007(3)$ as well as our six-loop resummation in Ref.\cite{abo-expon} that gives $\alpha=-0.00886$, one can easily realize the discrepancy between expected and so far calculated results from resummation of the $\varepsilon$-expansion of RG functions.   Even the resummation of the seven-loop $g$-expansion gives the result  $\alpha=-0.00859$ which in turn is still far away from the expected result. In view of our seven-loop result $\alpha=-0.0123(11)$ in this work and the mentioned previous results, one can claim that  the resummation of the seven-loop $\varepsilon$-expansion in this work is more than important.  

While the predictions of the renormalization group at fixed dimensions gives accurate results in three-dimensions \cite{zin-exp,Kleinert-Borel}, the story is different for the two dimensional cases. In two dimensions, the renormalization group at fixed dimensions gives inaccurate results especially for the  critical exponents of small values \cite{Borelg2,sokolov}. The reason behind this is the nonanaliticity of the $\beta$-function at the fixed point \cite{ON17,sokolov,plesito2}. The $\varepsilon$-expansion on the other hand might not suffer from this problem \cite{analytic}. We tested our resummation results in two dimensions and found an overall improvements to our  six-loop resummation results  in Ref.\cite{abo-expon}.  

Our algorithm while simple gives astonishing results for the critical exponents  which are competitive to the results from more sophisticated resummation algorithms, numerical methods as will as conformal field theory. This puts it among the preferred resummation algorithms applied to different problems in physics. A note to be mentioned here is that this work (up to the best of knowledge) represents the first resummation results for the $\varepsilon^7$ series in literature.


\begin{thebibliography}{99}                                                                                                


\bibitem {zinjustin}J. Zinn-Justin, Quantum Field Theory and Critical
Phenomena, International Series of Monographs on Physics Vol. 113, 4th ed.
(Clarendon Press, Oxford, 2002).

\bibitem {zin-borel}Jean Zinn-Justin and Ulrich D. Jentschura, J. Phys. A:
Math. Theor. 43, (2010).

\bibitem {Berzin}E. Br$\acute{e}$zin and G. Parisi, J. Stat. Phys. 19, 3 (1978).

\bibitem {Kleinert-Borel}H. Kleinert and V. Schulte-Frohlinde, Critical
Properties of $\phi^{4}$-Theories, World Scientific, Singapore) (2001).

\bibitem {kleinert}H. Kleinert, S. Thoms and W. Janke, Phys. Rev. A
\textbf{55}, 915 (1997).

\bibitem {kleinert2}Florian Jasch and Hagen Kleinert, J. Math. Phys. 42, 1 ( 2001).

\bibitem {zin-cr}J Zinn-Justin, Phys. Rep. 344, Issue 4-6, 159-178, April (2001).

\bibitem {Eta4}S.A. Antonenko and A.I. Sokolov, Phys. Rev. E 51, 1894 (1995).

\bibitem {Guillou}J. C. Le Guillou and J. Zinn-Justin, Phys. Rev. Lett. 39, 95 (1977).

\bibitem {Kleinert5L}H. Kleinert, J. Neu, V. Schulte-Frohlinde, K. G.
Chetyrkin and S. A. Larin, Phys. Lett. B 272, 39 (1991); Erratum, Phys. Lett.
B 319, 545(E) (1993).

\bibitem {pelsito}Andrea Pelissetto and Ettore Vicari, Phys. Rept.368:549-727 (2002).

\bibitem {MC10}M. Hasenbusch, Phys.Rev. B 82, 174433 (2010).

\bibitem {MC11}M. Hasenbusch and E. Vicari, Phys. Rev. B 84, 125136 (2011).

\bibitem {nuN0E}N. Clisby, J. Phys. A50, 264003 (2017).

\bibitem {nuN0}N. Clisby and B. Dunweg, Phys. Rev. E 94, 052102 (2016).

\bibitem {MCN2}M. Campostrini, M. Hasenbusch, A. Pelissetto, and E. Vicari,
Phys. Rev. B 74, 144506 (2006).

\bibitem {MC01}M. Hasenbusch J. Phys. A 34, 8221 (2001).

\bibitem {MC02}M. Campostrini, M. Hasenbusch, A. Pelissetto, P. Rossi, and E.
Vicari, Phys. Rev. B 65, 144520 (2002).

\bibitem {MC16}Ding, C., Blote, H.W.J. and Deng,Y. , Physical Review B 94,
104402 (2016).

\bibitem {MC19a} Wanwan Xu, Yanan Sun, Jian-Ping Lv, and Youjin Deng, Phys. Rev.
B 100, 064525 (2019).
\bibitem {MC19} Martin Hasenbusch, Phys. Rev. B 100, 224517 (2019).
\bibitem {Bstrab}Filip Kos, David Poland and David Simmons-Duffin, JHEP 06,091 (2014).

\bibitem {Bstrab5} Filip Kos, David Poland and David Simmons-Duffin, JHEP
11,106 (2015).

\bibitem {Bstrab2}Sheer El-Showk, Miguel F. Paulos, David Poland, Slava
Rychkov, David Simmons-Duffin and Alessandro Vichi, J. Stat.Phys.157:869-914 (2014)

\bibitem {Bstrab3}A. C. Echeverri, B. von Harling, and M. Serone, JHEP 09, 097 (2016).

\bibitem {Bstrab4}Filip Kos, David Poland, David Simmons-Duffin and Alessandro
Vichi, JHEP. 08, 036 (2016).

\bibitem {BstrabN0}Hirohiko Shimada and Shinobu Hikami, J. Stat. Phys.
165,1006 (2016).

\bibitem {NPRG}Gonzalo De Polsi, Ivan Balog, Matthieu Tissier and
Nicolás Wschebor, Phys. Rev. E 101, 042113 (2020).

\bibitem {zin-exp}R. Guida and J. Zinn-Justin, J.Phys. A 31, 8103 (1998). 
\bibitem {dispute} Shai M. Chester, Walter Landry, Junyu Liu, David Poland, David Simmons-Duffin, Ning Su and Alessandro Vichi,arXiv:1912.03324.
\bibitem {dispute1} Slava Rychkov, Journal Club for Condensed Matter Physics, :  \url{https://doi.org/10.36471/JCCM_January_2020_02}.

\bibitem {7L}Oliver Schnetz, Phys. Rev. D 97, 085018 (2018); Maple package
HyperlogProcedrues, which is available on the Oliver Schnetz's homepage https://www.math.fau.de/person/oliver-schnetz/
\bibitem {ON17}Mikhail V. Kompaniets and Erik Panzer, Phys.Rev. D.96, 036016 (2017).

\bibitem {Prd-GF}H\'{e}ctor Mera, Thomas G. Pedersen, and Branislav K.
Nikoli\'{c}, Phys.Rev. D.97.105027 (2018).
\bibitem {Prl}H\'{e}ctor Mera, Thomas G. Pedersen, and Branislav K.
Nikoli\'{c}, Phys. Rev. Let. 115, 143001 (2015).
\bibitem {cut}Thomas Garm Pedersen, H\'{e}ctor Mera and Branislav K.
Nikoli\'{}c, Phys.Rev. A 93, 013409 (2016).
\bibitem {error} B. Delamotte,  M. Dudka,  Yu. Holovatch,  and D. Mouhanna, Phys. Rev. B 82, 104432 (2010).
\bibitem {error1} B. Delamotte, Yu. Holovatch, D. Ivaneyko,  D.  Mouhanna,  and M. Tissier, J. Stat. Mech.: Theory Exp. (2008).
\bibitem {error2} A.  I.  Mudrov  and  K.  B.  Varnashev,  Phys.  Rev.  E   58,  5371(1998).
\bibitem {error3} J.  C.  Le  Guillou  and  J.  Zinn-Justin,  Phys.  Rev.  B   21,  3976(1980).
\bibitem {Borelg2}J.-P. Eckmann, J. Magnen, and R. S\'{e}n\'{e}or, Commun.
Math. Phys. 39, 251 (1975).
\bibitem {plesito2}A. Pelissetto and E. Vicari, Nucl. Phys. B 519 (3), 626 (1998).
\bibitem {sokolov}E. V. Orlov and A. I. Sokolov, Fizika Tverdogo Tela 42, 2087  
(2000) (Physics of the Solid State 42, 2151 (2000)).
\bibitem {analytic} L. Schafer, Phys. Rev. E 50,3517 (1994).
\bibitem {abo-expon} Abouzeid M. Shalaby, Phys. Rev. D 101, 105006 (2020).
\bibitem {Abo-large}Abouzeid Shalaby, Weak-Coupling, Strong-Coupling and
Large-Order Parametrization of the Hypergeometric-Meijer Approximants, arXiv:2002.05110.
\bibitem {alphaxy}J. A. Lipa, J. A. Nissen, D. A. Stricker, D. R. Swanson, and T. C. P. Chui, Phys. Rev. B 68, 174518 (2003).
\bibitem {HTF}Harry Bateman, HIGHER TRANSCENDENTAL FUNCTIONS, Volume I,
McGRAW-HILL BOOK COMPANY, INC. (1953).
\bibitem {epsilon7} Thomas A. Ryttov, JHEP 04,072 (2020).
\bibitem {nikel}B. G. Nickel, Physica A (Amsterdam) 117, 189 (1981).
\bibitem {exact}B. Nienhuis,  Phys. Rev. Lett. 49, 1062 (1982).

\bibitem {exact2}S. Caracciolo, A. J. Guttmann, I. Jensen, A. Pelissetto, A.
N. Rogers, and A. D. Sokal, J. Stat. Phys. 120, 1037 (2005).

\bibitem {Onsa}L. Onsager,  Phys. Rev.65, 117 (1944).

\bibitem {wn1}P. Calabrese, M. Caselle, A. Celi, A. Pelissetto, and E.
Vicari,  J. Phys. A 33, 8155 (2000).
\bibitem {Fractal} 	Mikhail Kompaniets and Kay Joerg Wiese, Phys. Rev. E 101, 012104 (2020).
\end{thebibliography}
\end{document}